\documentclass[conference]{IEEEtran}
\usepackage{cite}
\usepackage{amsmath,amssymb,amsfonts}
\usepackage{algorithmic}
\usepackage{graphicx}
\usepackage{textcomp}
\usepackage{xcolor}
\usepackage{hyperref}
\usepackage[braket, qm]{qcircuit}
\usepackage{todonotes}
\usepackage{dirtytalk}
\usepackage[shortlabels]{enumitem}

\def\BibTeX{{\rm B\kern-.05em{\sc i\kern-.025em b}\kern-.08em
    T\kern-.1667em\lower.7ex\hbox{E}\kern-.125emX}}
\begin{document}

\title{Simple Quantum State Encodings\\for Hybrid Programming of Quantum Simulators}

\author{\IEEEauthorblockN{Thomas Gabor, Marian Lingsch Rosenfeld, Claudia Linnhoff-Popien}
\IEEEauthorblockA{\textit{LMU Munich}
}}

\newcommand\copyrighttext{%
  \footnotesize This is a pre-print of an article published in \href{https://qsa-workshop.github.io/qsa2022/}{QSA 2022}.
  }
\newcommand\copyrightnotice{%
\begin{tikzpicture}[remember picture,overlay]
\node[anchor=south,yshift=10pt,xshift=10pt] at (current page.south) {\fbox{\parbox{\textwidth}{\copyrighttext}}};
\end{tikzpicture}%
}

\maketitle

\begin{abstract}
Especially sparse quantum states can be efficiently encoded with simple classical data structures. We show the admissibility of using a classical database to encode quantum states for a few practical examples and argue in favor of further optimizations for quantum simulation targeting simpler, only ``semi-quantum'' circuits.
\end{abstract}


\copyrightnotice{}

\section{Introduction}

As of today, most sufficiently complex quantum algorithms are not actually run on quantum hardware. Instead, quantum simulators have been known to offer more qubit space and even faster computation times per financial resource spent -- and most obviously: fully controllable noise~\cite{wille2019ibm}. As literature proposes many different approaches and suggestions on how to integrate future quantum hardware with classical computation infrastructure, integrating inherently classical quantum simulators with classical systems should be much easier. However, it is clear why the developers of quantum simulators may opt to more closely mimic the behavior of actual quantum computers in any aspect. Still, in this paper we want to consider using the classical crutch to our advantage while we still depend on it anyway. Effectively, we want to achieve two co-dependent goals: \begin{enumerate}[(A)]\item Write quantum circuits that incorporate classical operations to a larger extend than would be viable on a real quantum computer. \item Simulate circuits producing very simple, near-classical quantum states efficiently within a quantum simulator that runs on classical hardware anyway\end{enumerate}
In this paper, we show almost trivial examples for (A) and showcase how the simple approach of encoding quantum states as classical data structures as long as possible enables us to realize (B) in this context. As such, this paper should be read as a motivation to explore quantum algorithms beyond their current physically realizable platform. Why should that be useful? The success of quantum hardware will depend on quantum software and effective quantum software is in desperate need of new algorithms and new methods that yield algorithms with some kind of quantum advantage~\cite{gabor2020holy}. We reckon that finding such algorithms is made easier by expanding the playground and providing programming paradigms closer to what most non-quantum programmers are used to. As of today, whether we like it or not, the question which hardware platform our quantum algorithms will eventually be run on is secondary.

\section{Quantum State Encodings}

Qiskit is one of the most popular toolkits used to develop algorithms for gate-based quantum computers~\cite{wille2019ibm}. As such, it comes with an integrated simulator for quantum circuits that allows to test them on classical hardware. Our argument is motivated by the observation that for certain kind of quantum states very simple quantum state encodings perform much better at simulation than the state-of-the-art Qiskit simulator. We quantify this performance in Section~\ref{sec:experiments}.

What we mean by ``very simple state encodings'' is using the most obvious classical data structures to encode quantum states. For the purpose of this paper we compare the following approaches empirically:

\begin{enumerate}[(a)]
\item \textbf{Array.} We encode the entire state as a standard Numpy array containing all classical states occurring within the superposition and their respective amplitudes. Note that we only include states with non-zero amplitudes this way. 
\item \textbf{Database.} Similar to Array, we save all classical states with non-zero amplitude within the quantum state. In this case, we use an \texttt{sqlite3} database to do so.
\item \textbf{Qiskit.} As a benchmark, we perform all experiments using the state-of-the-art quantum simulator provided with Qiskit.
\item \textbf{Mixed.} We provide a simple heuristic to decide between using the Qiskit simulator and the Database state encoding on a case-by-case basis, i.e., if less than $\frac{2}{3}$ of the qubits have a Hadamard gate applied to them somewhere within the circuit, the Database encoding is used, and if not, the standard Qiskit simulator is used.
\end{enumerate}

The main idea behind Array and Database encoding is enabling efficient representation of very sparse quantum states. As most quantum algorithms try to avoid these states (as they do not exploit the capabilities of quantum hardware to the fullest extent), it is not surprising that standard quantum simulators like the one provided by Qiskit are not entirely optimized for these scenarios. In fact, using special encodings for rather sparse quantum states is not a new idea~\cite{vidal2003efficient}. However, we show that a state-of-the-art simulator like Qiskit (as future work should naturally also consider more quantum simulators) has a hard time recognizing near-classical states and optimizing its treatment of them. While at first it may seem paradoxical to use a quantum simulator to simulate almost or even fully classical algorithms, we argue that having an all-in-one simulation framework opens up a new process for hybrid programming, i.e., using quantum and classical parts within the same algorithm, where we can transfer classical programs identically to a quantum simulation framework to then embed quantum programming parts wherever they seem beneficial. Thus, in contrast to most work on encoding quantum states for quantum simulators (cf.~\cite{jozsa2006simulation,arad2010quantum,khammassi2017qx}, e.g.) we do not aim to improve classical encodings for standard quantum algorithms (although that is a valuable challenge, of course) but to reduce the overhead for already quasi-classical states within the simulator.

In the following experiments, we also present another twist on Database encoding called \emph{state drop}: We limit our database size to $1000$ entries by simply removing the quantum states with the smallest amplitude after each quantum (i.e., in our case Hadamard) operation. This approach was inspired by the technique of approximate approximation~\cite{sax2020approximate}, which works surprisingly well for quantum annealing. Our results are more expected and the error rate deteriorates quickly as our empirical results show. Still, for some experiments we see a small middle ground where computation times can be improved by admitting this approximation, perhaps making it worthwhile for further investigation.

\section{Experiments}
\label{sec:experiments}

To test the quantum state encodings described in the previous section, we execute various quantum algorithms. All code can be found in the accompanying  repository\footnote{\texttt{\url{https://github.com/marian-lingsch/qc-simulators}}}.

\subsection{Generate Superposition}

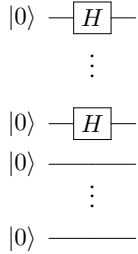
\begin{figure}
\centering
\resizebox{0.07\textwidth}{!}{
\Qcircuit @C=1em @R=1em {
   \lstick{\ket{0}} & \gate{H} & \qw \\
   & \vdots & \\
   &  & \\
   \lstick{\ket{0}} & \gate{H} & \qw \\
   \lstick{\ket{0}} & \qw & \qw \\
   & \vdots & \\
   &  & \\
   \lstick{\ket{0}} & \qw & \qw \\
}
}
\caption{Quantum circuit for the ``generate superposition'' benchmark.}
\label{fig:circuit-superposition}
\end{figure}

A Hadamard gate is applied to qubits $q_i$ for $i\in\{1, ..., r\}$, where for benchmarking purposes $r$ varied between $0$ and $n$ with $n$ being the total amount of qubits. See Figure~\ref{fig:circuit-superposition} for a depiction of the circuit.

\begin{figure}
    \centering
    \begin{minipage}{0.45\textwidth}
        \centering
        \includegraphics[width=0.75\textwidth]{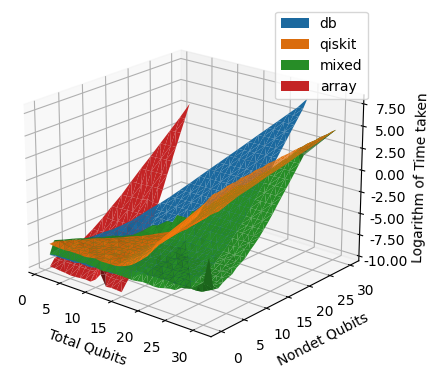}
    \end{minipage}\hfill
    \begin{minipage}{0.45\textwidth}
        \centering
        \includegraphics[width=0.75\textwidth]{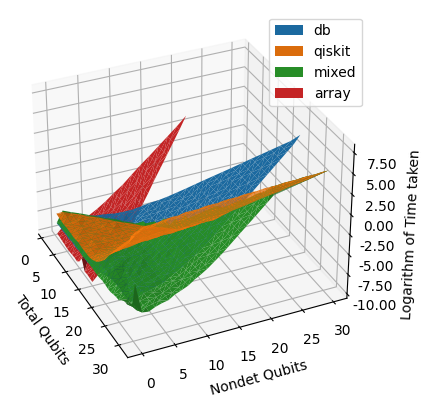}
    \end{minipage}
    \caption{Performance analysis for the ``generate superposition'' benchmark. Identical 3D plot from two perspectives.}
    \label{fig:analysis-superposition}
\end{figure}

\begin{figure}
        \centering
        \includegraphics[width=0.4\textwidth]{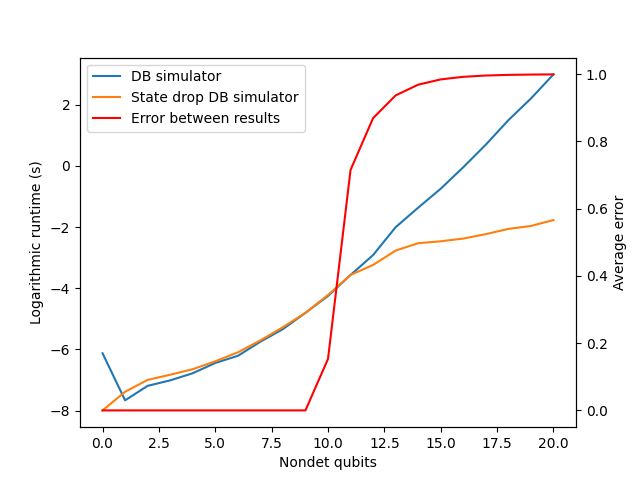}
    \caption{State drop performance for the ``generate superposition'' benchmark}
    \label{fig:drop-superposition}
\end{figure}

We execute this benchmark using all quantum state encodings described in the previous section. Note that we vary the overall size of the benchmark by simply introducing more qubits in total, i.e., increasing $n$. We also vary the ``quantum-ness'' of the circuit by applying the Hadamard gate only to some (i.e., an amount of $r$) of the involved qubits (shown on the ``Nondet qubits'' axis in Figure~\ref{fig:analysis-superposition}) as only these qubits' values cannot be determined classically. The Z axis shows the execution time in seconds on a log scale. When running the experiments on our test machine was no longer feasible in regard to computational resources, we cut off the experiments at the largest possible number of total qubits. Note that with respect to the base surface (i.e., ``Total qubits''$\times$``Nondet qubits'') all plotted surfaces occupy the upper triangle, i.e., ``Nondet qubits'' can never be larger than ``Total qubits''.

It can be seen that the runtime of the database simulator is only dependent on the amount of \say{Nondet Qubits} or $r$, while the runtime of the Qiskit simulator depends on the \say{Total Qubits} or $n$, even when these add no real complexity to the problem. We observe that for few \say{Nondet Qubits} the Database simulator performs better than the Qiskit simulator, while the Qiskit simulator is better than the Database simulator when the amount of \say{Nondet Qubits} is close to the amount of \say{Total Qubits}; in fact Qiskit is the only computationally feasible encoding for more than $27$ ``Nondet qubits''. The Mixed simulator allows us to have the best of both worlds in all instances. It can be seen that the Array simulator performs much worse than the Database simulator, except for very small inputs. For this reason, it was not considered in further experiments.

Figure~\ref{fig:drop-superposition} shows the results for the state drop technique. We fixed the total number of qubits to $20$ and compared computation times and produced error for both the standard Database simulator as well as Database with state drop. Note that the pure Database error rate is always $0$ so it is not plotted separately. In this scenario, we see that computation times of state drop compared to the full Database approach only improve in a regime where error rates have already reached infeasible heights.

\subsection{Addition}
\label{ssec:addition}

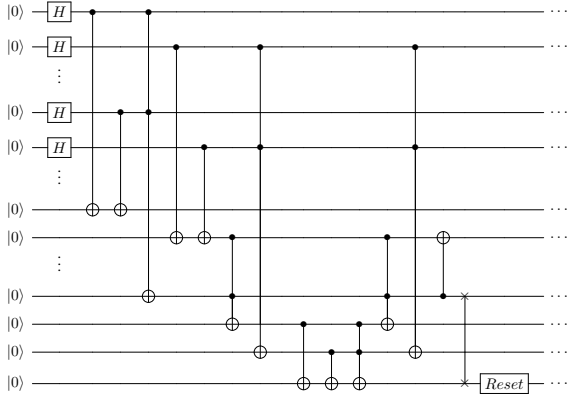
\begin{figure}
\centering
\resizebox{0.39\textwidth}{!}{
\Qcircuit @C=1em @R=1em {
   \lstick{\ket{0}}  & \gate{H} & \ctrl{8} & \qw & \ctrl{12} & \qw & \qw & \qw & \qw &  \qw & \qw & \qw & \qw & \qw & \qw & \qw & \qw & \qw & \qw & \cdots \\
   \lstick{\ket{0}}  & \gate{H} & \qw & \qw  & \qw & \ctrl{8} & \qw & \qw & \ctrl{13} & \qw & \qw & \qw & \qw & \qw & \ctrl{13} & \qw & \qw & \qw & \qw & \cdots \\
   & \vdots & & \\
   &  & &  \\
   \lstick{\ket{0}} & \gate{H} & \qw & \ctrl{4} & \ctrl{8} & \qw & \qw &  \qw  & \qw & \qw & \qw & \qw & \qw & \qw & \qw & \qw & \qw & \qw & \qw & \cdots \\
   \lstick{\ket{0}}  & \gate{H} & \qw & \qw & \qw & \qw & \ctrl{4} & \qw & \ctrl{8} & \qw & \qw & \qw & \qw & \qw & \ctrl{9} & \qw & \qw & \qw & \qw & \cdots \\
   & \vdots & & \\
   &  &  & \\
   \lstick{\ket{0}} & \qw & \targ & \targ & \qw & \qw & \qw & \qw & \qw & \qw & \qw & \qw & \qw & \qw & \qw & \qw & \qw & \qw & \qw & \cdots \\
   \lstick{\ket{0}} & \qw & \qw & \qw & \qw & \targ & \targ & \ctrl{4} & \qw & \qw & \qw & \qw & \qw & \ctrl{4} & \qw & \targ & \qw & \qw & \qw & \cdots \\
   & \vdots & & \\
   &  & & \\
   \lstick{\ket{0}} & \qw & \qw & \qw & \targ & \qw & \qw & \ctrl{1} & \qw & \qw & \qw & \qw & \qw & \ctrl{1} & \qw & \ctrl{-3}  & \qswap & \qw & \qw & \cdots \\
   \lstick{\ket{0}} & \qw & \qw & \qw & \qw & \qw & \qw & \targ & \qw & \qw & \ctrl{2} & \qw &  \ctrl{2} & \targ & \qw & \qw & \qw \qwx & \qw & \qw & \cdots \\
   \lstick{\ket{0}} & \qw & \qw & \qw & \qw & \qw & \qw & \qw & \targ & \qw &  \qw & \ctrl{1} &  \ctrl{1} & \qw & \targ & \qw & \qw \qwx & \qw & \qw & \cdots \\
   \lstick{\ket{0}} & \qw & \qw & \qw & \qw & \qw & \qw & \qw & \qw & \qw & \targ & \targ & \targ & \qw & \qw & \qw & \qswap \qwx & \gate{Reset} & \qw & \cdots \\
}
}
\caption{Quantum circuit for the ``addition'' benchmark.}
\label{fig:circuit-addition}
\end{figure}

We add two input variables with $k$ bits each. For benchmarking purposes, we apply Hadamard gates to $r$, $r \in \{0, ..., 2k-1\}$, of these qubits. Qubits $q_0$ through $q_{2k-1}$ represent the input variables and $q_{2k}$ through $q_{3k - 1}$ represent the output variable while $q_{3k}$ through $q_{3k + 4}$ are ancillary qubits. The total amount of qubits is $3k + 5$. Figure~\ref{fig:circuit-addition} depicts the whole circuit.

\begin{figure}
    \centering
    \begin{minipage}{0.45\textwidth}
        \centering
        \includegraphics[width=0.75\textwidth]{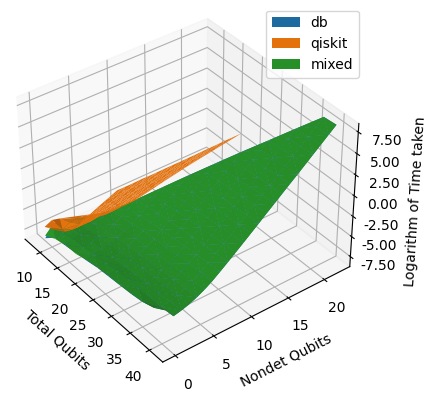}
    \end{minipage}\hfill
    \begin{minipage}{0.45\textwidth}
        \centering
        \includegraphics[width=0.70\textwidth]{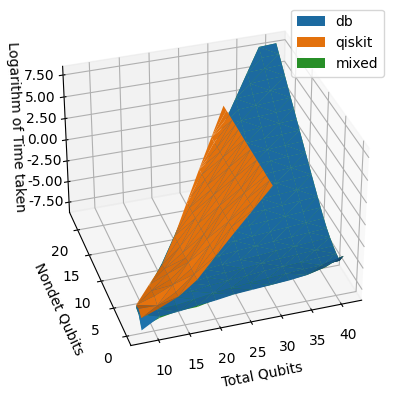}
    \end{minipage}
    \caption{Performance analysis for the ``addition'' benchmark. Identical 3D plot from two perspectives.}
    \label{fig:analysis-addition}
\end{figure}

\begin{figure}
        \centering
        \includegraphics[width=0.4\textwidth]{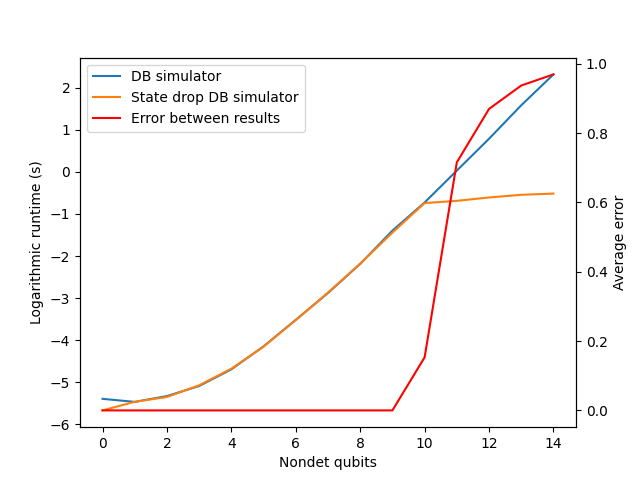}
    \caption{State drop performance for the ``addition'' benchmark}
    \label{fig:drop-addition}
\end{figure}

The analysis in Figure~\ref{fig:analysis-addition} shows that the Qiskit simulator performs worse than the Database simulator most of the time. This occurs because there are much fewer states with non-zero probability than the total possible amount, given the mostly classical nature of the circuit. The Mixed simulator always chooses to use the Database simulator and thus their runtimes are very similar. For more than $29$ total qubits, Qiskit could no longer be run practically on our machine while Database (and thus Mixed) still work up to $30$ total qubits.

Figure~\ref{fig:drop-addition} shows the state drop trade-off for $26$ total qubits. At $11$ ``Nondet qubits'' we already have a huge normalized error and errors quickly climb to maximum beyond that. However, compared to Figure~\ref{fig:drop-superposition}, there exists a trade-off at least, when with $11$, $12$, or $13$ ``Nondet qubits'' we save computation time (albeit by introducing great amounts of error), which could be further investigated in future work.

\subsection{Grover's Search}

\begin{figure}
\centering
\resizebox{0.25\textwidth}{!}{

\Qcircuit @C=1em @R=1em {
   \lstick{\ket{0}}  & \gate{H} & \multigate{8}{A} & \qw & \multigate{4}{D} & \qw & \multigate{4}{D} & \qw & \cdots \\
   \lstick{\ket{0}} & \gate{H} & \ghost{A} & \qw & \ghost{D} & \qw & \ghost{D} & \qw & \cdots \\
   & \vdots & &  & &  & &  & \\
   &  & &  & &  & &  & \\
   \lstick{\ket{0}} & \gate{H} & \ghost{A} & \qw & \ghost{D} &  \qw & \ghost{D} & \qw & \cdots \\
   \lstick{\ket{0}} & \qw & \ghost{A} & \multigate{3}{U} & \qw & \multigate{3}{U} & \qw & \qw & \cdots \\
   & \vdots & &  & &  & &  & \\
   &  & &  & &  & &  & \\
   \lstick{\ket{0}} & \qw & \ghost{A} & \ghost{U} & \qw & \ghost{U} & \qw & \qw & \cdots \\
}

}
\caption{Quantum circuit for the ``Grover's search'' benchmark}
\label{fig:circuit-grover}
\end{figure}
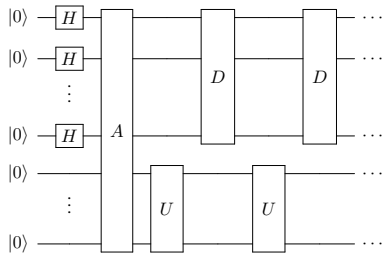

We implement Grover's algorithm using an oracle for $\mathbf{0}$, i.e., the state where all the numbers are zero. The circuit used to implement this algorithm is given in Figure~\ref{fig:circuit-grover}, where $U$ is the oracle, $A$ is the addition operator as described in Subsection~\ref{ssec:addition}, and $D$ is the diffusion operator.

For the Qiskit simulator the diffusion operator is implemented as a series of Hadamard and CNOT gates, while for the Database simulator we took a shortcut by implementing the effect of the diffusion operator within a single SQL query to the database to showcase the potential of classical approximation for practical simulation.

\begin{figure}
    \centering
    \includegraphics[width=0.4\textwidth]{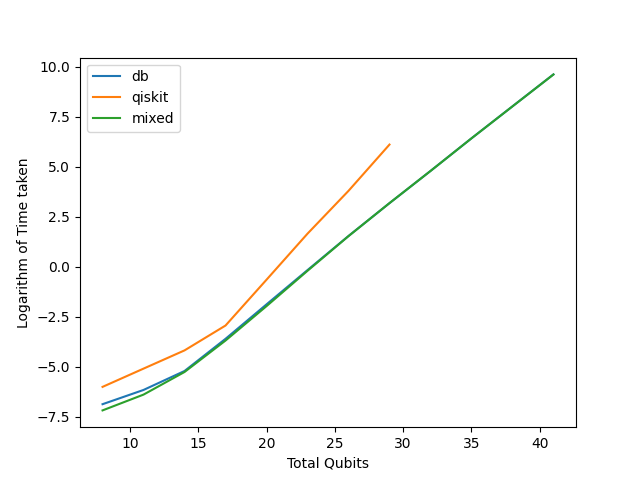}
   	\caption{Performance analysis for the ``Grover's search'' benchmark.}
   	\label{fig:analysis-grover}
\end{figure}

\begin{figure}
        \centering
        \includegraphics[width=0.4\textwidth]{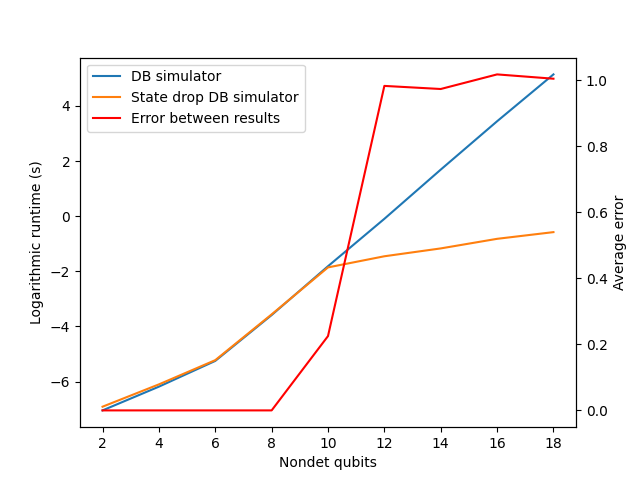}
    \caption{State drop performance for the ``Grover's search'' benchmark}
    \label{fig:drop-grover}
\end{figure}

Since all qubits in the search range need to be non-deterministic for Grover's algorithm, Figure~\ref{fig:analysis-grover} shows a two-dimensional plot this time. As the oracle implements addition, the Mixed simulator's behavior carries over from the previous benchmark (cf. Subsection~\ref{ssec:addition}) and it always resorts to the Database simulator, which performs better than the Qiskit simulator. It is interesting to observe that behavior even when the whole framework of Grover's search, one of the primary examples for quantum algorithms, is put around the benchmark of addition.

Figure~\ref{fig:drop-grover} shows a discouraging image where state drop errors rise early on and we again only see a small regime at $11$ ``Nondet qubits'' (i.e., where the ``Nondet qubits'' span over $2^{11} = 2048$ possible states) where we can achieve a computational benefit by accepting relatively large error rates.

\section{Conclusion}

We have analyzed four basic techniques for classically encoding quantum states and have seen that even very simple mechanisms can beat the state-of-the-art Qiskit simulator, when the circuit considered do not need to construct too complex quantum states. We have shown a very simple ``Mixed'' heuristic that can switch between these scenarios dynamically. Future work needs to improve that heuristic for more complex real-world scenarios and perhaps integrate many more means of quantum state simulation. Furthermore, the ``state drop'' technique was first evaluated with limited success. However, we reckon that more advanced heuristics might still be able to more precisely hit a sweet spot between computational resources and acceptable error that might just push classical quantum simulation a step further.

Overall, we think it is big challenge for the field to come up with a fully integrated simulator that employs various techniques to recognize the circuit it is given and that can be easily parametrized to yield an arbitrary error/time trade. We should also consider how much more complex algorithms might be split up and be simulated using a combination of various techniques. We suspect that such a generic platform for quantum software development will enable more complex applications for quantum computers by allowing developers to start out with classical algorithms as they would use on classical hardware and migrate gradually towards more usage of quantum states and thus quantum hardware. Such tools could lead to a truly hybrid workflow for programming quantum computers to be used within hybrid setups for real-world challenges.

\end{document}